\documentclass[conference]{IEEEtran} 
\usepackage{mathtools}
\usepackage{cite} 
\usepackage{graphics}
\usepackage{graphicx}
\usepackage[english]{babel}
\usepackage{amsmath}
\usepackage{amsthm}
\usepackage[table,xcdraw]{xcolor}
\usepackage{subfig}
\usepackage{enumitem}
\usepackage[linesnumbered, ruled]{algorithm2e}
\usepackage[noend]{algpseudocode}
\usepackage[font=footnotesize]{caption}
\usepackage{epsfig}
\usepackage{xfrac}
\usepackage{nicefrac}
\usepackage{psfrag}
\usepackage{times}
\usepackage{balance} 
\usepackage{amssymb}
\usepackage{amsfonts}
\usepackage{bm}
\usepackage[bottom]{footmisc}
\usepackage{mwe}
\usepackage{color}
\usepackage{tabularx}
\usepackage{booktabs,multirow}
\usepackage{verbatim}
\usepackage{hyperref}
\usepackage{multirow}
\usepackage{mwe}
\usepackage{booktabs,multirow}
\usepackage{tikz}
\usetikzlibrary{patterns,arrows}
\usepackage{physics}
\usepackage{verbatim}
\usepackage{wrapfig}
\usepackage{soul}
\usepackage[left=1.62cm,right=1.62cm,top=1.9cm]{geometry}

\usepackage{xcolor}

\usepackage{titlesec}

\DeclareMathOperator*{\argmax}{arg\,max}

\SetKwRepeat{Do}{do}{while}
\SetKwInOut{Input}{Input}
\SetKwInOut{Output}{Output}
\SetKw{KwBy}{by}

\usepackage{lipsum}

\usepackage{acronym}
\newacro{UE}{User Equipment}
\newacro{BS}{Base Station}
\newacro{PU}{Primary User}
\newacro{SU}{Secondary User}
\newacro{PBS}{Primary Base Station}
\newacro{SBS}{Secondary Base Station}
\newacro{SNR}{Signal-to-noise Ratio}
\newacro{SINR}{Signal-to-interference-plus-noise Ratio}
\newacro{CR}{Cognitive Radio}
\newacro{SS}{Spectrum Sensing}
\newacro{MPC}{Multipath Component}
\newacro{DoD}{Direction of Departure}
\newacro{DoA}{Direction of Arrival}
\newacro{ML}{Maximum Likelihood}
\newacro{Tx}{Transmitter}
\newacro{Rx}{Receiver}
\begin{document}

\title{\fontsize{21}{24}\selectfont Cognitive Radio for Asymmetric Cellular Downlink with Multi-User MIMO}
\author{ 
Omer Gokalp Serbetci, {\small {\em Student Member, IEEE}}, Lei Chu, {\small {\em Senior Member, IEEE}} and Andreas F. Molisch, {\small {\em Fellow, IEEE}} \\
\small Ming Hsieh Department of Electrical and Computer Engineering, University of Southern California,
\small Los Angeles, USA\\
\small \{serbetci, lc\_285, molisch\}@usc.edu
}

\maketitle
\begin{abstract} Cognitive radio (CR) is an important technique for improving spectral efficiency, letting a secondary system operate in a wireless spectrum when the primary system does not make use of it. While it has been widely explored over the past 25 years, many common assumptions are not aligned with the realities of 5G networks. In this paper, we consider the CR problem for the following setup: (i) infrastructure-based systems, where downlink transmissions might occur to \acp{Rx} whose positions are not, or not exactly, known; (ii) multi-beam antennas at both primary and secondary base stations. We formulate a detailed protocol to determine when secondary transmissions into different beam directions can interfere with primary users at {\em potential} locations and create probability-based interference rules. We then analyze the "catastrophic interference" probability and the "missed transmission opportunity" probability, as well as the achievable throughput, as a function of the transmit powers of the primary and secondary base stations and the sensing window of the secondary base station. Results can serve to more realistically assess the spectral efficiency gains in 5G infrastructure-based cognitive systems. 

\end{abstract}

\section{Introduction} \label{sec: intro}


As the demand for wireless services increases, new adaptive methods for utilizing the wireless spectrum become necessary. Traditionally, spectrum has been assigned to specific entities, such as network operators or defense agencies. However, newly released bands are often made available with the stipulation that (temporarily) unused frequency resources can be exploited by users who do not belong to the nominal "owner" of the spectrum. In particular, the recent establishment of the Citizen Broadband Radio Service by the Federal Communication Commission, in which new services such as 5G are allowed to operate in certain mid-band (3.5 GHz) bands with legacy users has added significant relevance to this topic both from the military and civilian/economic point of view \cite{FCC}.

The trend toward adaptive spectrum usage began more than 20 years ago under the name of \ac{CR} and has seen a surge of interest, e.g., \cite{biglieri2013principles,setoodeh2017fundamentals}. 
A cognitive radio system is characterized by the presence of a {\em primary} system, which includes infrastructure and user nodes, as well as a secondary system. The key requirement is that the activities of the secondary system do not (or only within specified limits) degrade the performance of the primary system. To ensure this, the secondary system must monitor the spectrum occupancy of the primary system and decide whether or not to transmit in order to avoid interference \cite[Chap. 26]{molisch2023wireless}.

Although there are thousands of papers on \ac{CR}, the current work focuses on a scenario that has received comparatively little attention despite being practically important: {\em \ac{CR} in an infrastructure-based system using directional transmissions, where downlink transmissions may occur to \acp{UE} whose locations are unknown or only partially known.} Conventional \ac{CR} approaches often assume that (i) nodes transmit in all directions and (ii) that primary nodes operate in a manner that allows the secondary system to infer the location of, or the propagation channel to, the potential victim (primary) \acp{Rx}, see also Sec. \ref{sec: system_model}. Neither of these assumptions typically holds in a 5G system. Regarding (i), the use of multiple antennas (and hence beamforming capabilities in various directions) is a mandatory part of the 3GPP NR standard \cite{dahlman20205g}. Regarding (ii), identifying the location (or channel) of the currently {\em actively receiving} \ac{PU}, i.e., the victim \ac{Rx}, can be difficult or even impossible. In the 5G standard, there are generally no uplink signals (i.e., from the \ac{PU} to the \ac{PBS}) that {\em precede} downlink transmission; see Sec. \ref{sec: algorithm} for details.

To overcome these limitations, this paper presents a detailed model for directional sensing and decision-making in the {\em downlink} of an {\em infrastructure-based} system (e.g., cellular 5G), where both the \ac{PBS} and \ac{SBS} use multi-beam (sectorized) antennas. While the channels between the \ac{PBS} and all {\em potential} \ac{PU} locations are assumed to be known (e.g., through training and feedback over time), the identities and locations of the {\em currently active} \acp{PU} (i.e., those receiving downlink transmissions) are unknown. The \ac{SBS} must therefore perform three tasks: (i) learn - through a process described in Sec. \ref{sec: algorithm} - whether transmission from each \ac{SBS} beam creates "significant" interference, defined via the \ac{SINR}, to each {\em potential} \ac{PU} location; (ii) sense, in each of its beams, the transmission from the \ac{PBS} to determine which \ac{PBS} beams are currently active - a task that can be formulated as a multi-binary hypothesis testing problem; and (iii) based on this information, determine which \ac{SBS} beams are permitted to transmit without exceeding a specified statistical interference threshold at the \acp{PU}.

The main contributions of this work are thus
\begin{itemize}
    \item Formulation of the cognitive problem for the downlink in the infrastructure-based primary and secondary system, undetectable location/channels of the \acp{PU}, and multi-beam \ac{PBS} and \ac{SBS}.
    \item Designing a novel downlink transmission decision rule for the SBS.
    \item Performing simulations based on realistic (ray-tracing) channel configurations, demonstrating the performance.
\end{itemize}
\section{Related Work}
Since \ac{SS} and \ac{CR} are well-established research areas, numerous aspects have been explored over the years, including sensing approaches and transmission decision criteria and algorithms, based on a variety of system models, see \cite{biglieri2013principles,setoodeh2017fundamentals} and references therein. 

Most prior works assume omnidirectional antennas and formulate sensing as a binary hypothesis test over the entire region. Directional setups, when explored, often focus on single-source localization \cite{cabric} or transmitter detection using switched-beam antennas in distributed networks \cite{kleber2021directivity}. Cooperative sensing has been studied in both distributed \cite{distributed} and centralized \cite{golmie} frameworks. Still, these typically assume full-duplex operation or synchronized Tx/Rx cycles that permit estimation of channels from secondary TXs to currently active primary \acp{Rx} — assumptions that do not hold in our system model. Additional studies explore directional sensing via stochastic geometry \cite{stoc_geom,directional_coverage}, capacity analysis for directional \ac{SU} networks \cite{yazdani}, and beam selection strategies to improve detection performance \cite{golmie}.

More related to our case is \cite{deepallocation}, which bypasses explicit sensing and directly allocates \ac{SBS} power based on sensing inputs, without modeling the \ac{PBS}. Key differences include: (i) lack of a communication model; (ii) knowledge of instantaneous PU activity via dedicated sensors, unlike our use of \ac{PBS} downlink signals; (iii) interpolation of sparse data versus our learned channel estimates; and (iv) a strict zero-interference constraint, while we allow controlled interference.

The most comparable work is \cite{awe}, which tackles the beam detection subtask but relies on supervised learning with labels from the primary system to the secondary system. In contrast, we assume that such collaboration is not allowed/feasible, as it requires changes in the primary system; rather, we aim to detect active \ac{PBS} beams solely from \ac{SBS} observations and select interference-aware \ac{SBS} beams to serve \acp{SU}. Thus, our work fundamentally differs in system modeling and operational assumptions.

\section{System Model} \label{sec: system_model}  
This section outlines the operation of the \ac{PBS} and \ac{SBS}, along with the associated signal and channel models. The PBS operates with a common slotted time structure consisting of $I$ equal-length intervals. During each interval $i \in \mathcal{I} \triangleq \{1, 2, \dots, I\}$, spanning $T_{i-1} \leq t \leq T_i$, the \ac{PBS} transmits packets to the \acp{PU}.  
We define the set of all possible \ac{UE} locations (either \acp{PU} or \acp{SU}) within the region of interest as $\mathcal{S} \triangleq \{\boldsymbol{z}_1, \boldsymbol{z}_2, \dots, \boldsymbol{z}_U\}$, where $\boldsymbol{z}_u \in \mathbb{R}^2$ denotes the coordinates of \ac{UE} $u \in \mathcal{U}$. The set of UE indices is denoted by $\mathcal{U} \triangleq \{1, 2, \dots, U\}$.
\subsection{Multi-Beam Configuration at the Base Stations}
We assume that the \ac{PBS} and \ac{SBS} have $B_{\rm PBS}$ and $B_{\rm SBS}$ beams, respectively, {dividing the azimuth plane into (potentially non-uniform) sectors} Each beam, indexed $k$, of the \acp{PBS} might be active (ON) or inactive (OFF) during interval $i$, as indicated by the  $b_{i,k}^{\rm PBS} \in \{0,1\}$; the set of activity indicators in slot $i$ is  $\mathcal{B}_i^{\rm PBS} \in \{0,1\}^{B_{\rm PBS}}$. 

\begin{figure}[h!]
    \centering
    \includegraphics[trim = {0, 2cm, 0, 0}, clip, width= 1\columnwidth]{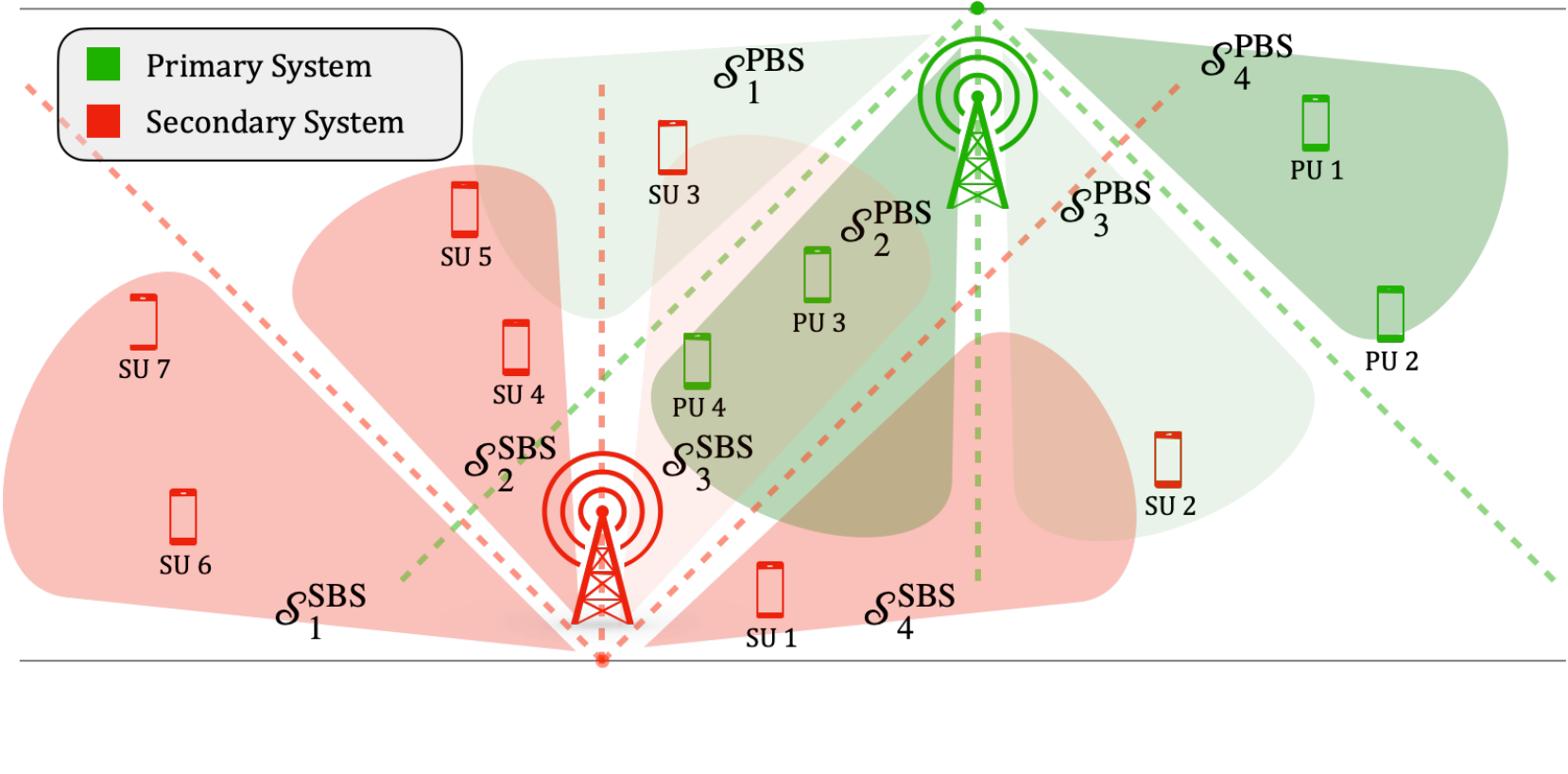}
    \caption{Illustration of primary and secondary beams with scattered \acp{PU} and \acp{SU}; inactive beams are shown with light shading.}
    \label{fig:system_model}
\end{figure}

The \ac{SBS} adopts the same basic time structure as the \ac{PBS}, but each interval $i$ is divided into two sub-intervals: sensing ($T_{i-1} \leq t \leq T_i^{\text{sensing}}$) and transmission ($T_i^{\text{sensing}} \leq t \leq T_{i}$). During the $i^{\text{th}}$ sensing phase, the \ac{SBS} collects $N$ samples indexed by $n \in \{1, \dots, N\}$ in each beam, thereby obtaining observations of the \ac{PBS} signals. We assume that the \ac{SBS} can obtain samples on all its beams simultaneously (i.e., no beam sweeping is required). 
Based on these observations, the \ac{SBS} determines the set of beams it uses for transmission in the $i^{\text{th}}$ interval, similarly represented as $\mathcal{B}_i^{\rm SBS} = \{b_{i,l}^{\rm SBS}\}^{B_{\rm SBS}}_{l = 1}$. The \ac{SBS} knows $B_{\rm PBS}$ and the $T_{i}$; for notational simplicity, we further assume that both the primary and secondary systems operate over the same known bandwidth $BW$, consisting of $F$ equally spaced frequencies $ \mathcal{F} \triangleq \{f_j\}_{j = 1}^F$. All these parameters can be obtained from standards or equipment specifications. However, we stress that (i) the \ac{SBS} does not know with which \acp{PU} the \ac{PBS} communicates, and (ii) it must infer from its sensing results which \ac{PBS} sectors are active.
Fig.~(\ref{fig:system_model}) illustrates an example configuration, showing beam orientations for both \ac{PBS} and \ac{SBS}, along with example \acp{PU} and \acp{SU} in the environment.

The $n^{\text{th}}$ signal sample received in the $l$-th beam of the \ac{SBS}, corresponding to the transmission from the \ac{PBS} on frequency $f_j$ during the $i^{\text{th}}$ sensing interval, is denoted by $r^{\rm SBS}_{i, j, l, n}$ is 
\begin{equation} \label{eq: sbs_rx_signal}
    r^{\rm SBS}_{i, j, l, n}  = \sum_{k = 1}^{B_{\rm PBS}}b_{i,k}^{\rm PBS} \cdot h^{\rm PBS\rightarrow \rm SBS}_{j, k, l} \cdot s^{\rm PBS}_{i, j, k, n}  + w_{i, j, l, n} ~.
\end{equation}
The signal transmitted on the $k$-th beam at frequency $f_j$ is denoted as $s^{\rm PBS}_{i,j,k,n}$ and is modeled as Gaussian random variable$ \sim \mathcal{CN}(0, \sigma^2_{\rm PBS})$ since both the payload data and the modulation format are unknown to the \ac{SBS}. The complex channel transfer function at frequency $f_j$ from the  $k^{\text{th}}$ \ac{PBS} beam to the $l^{\text{th}}$ \ac{SBS} beam is denoted as $h^{\rm PBS \rightarrow SBS}_{j,k,l} \in \mathbb{C}$ and assumed to be constant over the duration of a packet. Each received sample is also corrupted by additive complex Gaussian noise $w_{i,j,l,n} \sim \mathcal{CN}(0, \sigma^2_{\text{noise}})$, where noise power is assumed constant across frequency, beam, and time.

The complex channel gain can be described as the sum of $M$ weighted contributions of the \acp{MPC} 
\begin{equation} \label{eq: bs_channel}
h_{j, k, l}^{\rm PBS \rightarrow SBS} = \sum_{m = 1}^M |\alpha_{m}|e^{\text{i} \phi_{m}}e^{- \text{i}2\pi\tau_{m}f_j}\beta_{k}(\Omega_m)\beta_{l}(\Psi_m) 
\end{equation}
Here, the complex gain, phase, delay, direction of departure (DoD), and direction of arrival (DoA) of the \( m^{\text{th}} \) \ac{MPC} are denoted by \( \alpha_m \), \( \phi_m \), \( \tau_m \), \( \Omega_m \), and \( \Psi_m \), respectively. Moreover, the antenna pattern corresponding to the \( k^{\text{th}} \) beam of the \ac{PBS} is represented by 
\begin{equation}
\beta_{k}(\Omega_m) = 
\sum_{n=1}^{B_{\text{PBS}}} e^{jn \pi \left( \cos\Omega_m - \cos\Omega_k \right)}
\quad,
\end{equation}
where \( \Omega_k \) is the steering angle for sector \( k \). The pattern $\beta_l(\Psi_m)$ for the \( l^{\text{th}} \) beam of the \ac{SBS} is defined similarly.

We partition the location set $\mathcal{S}$ into mutually exclusive subsets corresponding to the \ac{PBS} beam sectors as $\mathcal{S} = \bigcup_{k = 1}^{B_{\rm PBS}} \mathcal{S}_k^{\rm PBS}$, where $\mathcal{S}_k^{\rm PBS} \cap \mathcal{S}_l^{\rm PBS} = \emptyset$ for all $k \neq l$. The corresponding \ac{PU} index set is similarly partitioned as $\mathcal{U} = \bigcup_{k = 1}^{B_{\rm PBS}} \mathcal{U}_k^{\rm PBS}$. The association between a \ac{UE} and a beam of the \ac{BS} is based on the small-scale-averaged path gain. In line-of-sight (LOS) scenarios, this association may admit a geometric interpretation, though this is not strictly required. The beam associated with the UE located at $\boldsymbol{z}_u$ with respect to the \ac{PBS} is defined as 
\begin{equation}
c_u^{\rm PBS} \triangleq \argmax_k \frac{1}{F} \sum_{j = 1}^F |h_{j,k,u}^{\rm BS \rightarrow UE}|^2.
\end{equation}
The corresponding beam associations with respect to the \ac{PBS} are denoted by $c_u^{\rm PBS}$. Accordingly, $c_u^{\rm PU} = k$ for all $u \in \mathcal{U}_k^{\rm PBS}$.

For ease of notation, we discuss the case of a single \ac{PBS}. Generalization to multiple \acp{PBS} is straightforward as it generates a larger set of \ac{PBS} sectors (and similarly for SBS), and the same data collection and transmission decision rules can be applied; however, it increases the numerical complexity of obtaining solutions. 

\subsection{Interference Estimate in Asymmetrical Downlink} \label{sec:interference-estimate}
To assess potential performance degradation at the \acp{PU}, we model the received signals as a function of the \ac{SBS} operation. During $T_i^{\text{sensing}} \leq t \leq T_i$, the \ac{SBS} uses the set of beams $\mathcal{B}_i^{\rm SBS}$ (selection of this set is discussed in Sec.~\ref{sec: algorithm}). The received signal at the \ac{PU} is modeled as the sum of the signals from the active \ac{PBS} and \ac{SBS} beams, each weighted by its respective channel. The received sample is further corrupted by additive complex Gaussian noise. The signal transmitted by the \ac{SBS}, similar to that of the \ac{PBS}, is modeled as a zero-mean complex Gaussian random variable with power \( \sigma^2_{\rm SBS} \).
The channel at $f_j$ between the \( k^{\rm th} \) beam of a base station and a \ac{UE} located at position \( \boldsymbol{z}_u \) is 
\begin{equation} \label{eq: bs_ue_channel}
h_{j, k, u}^{\rm BS \rightarrow UE} = \sum_{m = 1}^M |\alpha_{u, m}|e^{\text{i} \varphi_{u, m}}e^{- \text{i}2\pi\tau_{u, m}f_j}\beta_{k}^{\rm BS}(\Omega_{u,m}) ~.
\end{equation}
The definitions of the \acp{MPC} are analogous in Eq. (\ref{eq: bs_channel}). The only difference is the absence of the antenna gain term corresponding to the direction of arrival, since the \acp{UE} are assumed to have omnidirectional antennas. Associations between a \ac{PU} and an \ac{SBS} sector are denoted by  $c_u^{\rm SBS}$, analogously to the association between \ac{PU} and a \ac{PBS} sector.  
The SINR for \ac{PU} $u$ at location $\boldsymbol{z}_u$ during interval $i$ is defined as $\text{SINR}_{i,u} \triangleq \frac{P_{i,u}^{\rm signal}}{P_{i,u}^{\rm int} + P_{i,u}^{\rm noise}}$. The received signal power from the associated \ac{PBS} beam is given by 
\begin{equation}
P_{i,u}^{\rm signal} \triangleq b_{i, c_u^{\rm PBS}}^{\rm PBS} \, \sigma^2_{\rm PBS} \, \frac{1}{F} \sum_{j = 1}^{F} | h_{j, c_u^{\rm PBS}, u}^{\rm PBS \rightarrow PU} |^2.
\end{equation}
The total interference power is decomposed as $P_{i,u}^{\rm int} \triangleq P_{i,u}^{\rm Pint} + P_{i,u}^{\rm Sint}$, where $P_{i,u}^{\rm Pint}$ accounts for inter-beam interference from the \ac{PBS} 
\begin{equation} \label{eq: pbs_interference}
P_{i, u}^{\rm Pint} \triangleq \sigma^2_{\rm PBS} \sum\limits_{\substack {k = 1 \\ k \neq c_u^{\rm PBS}}}^{B_{\rm PBS}}b_{i,k}^{\rm PBS} \frac{1}{F}\sum_{j = 1}^F \left |h_{j, k, u}^{\rm PBS\rightarrow \rm PU} \right |^2
\end{equation}
while the interference from \ac{SBS} transmissions is 
\begin{equation}
  P_{i,u}^{\rm Sint} \triangleq \sigma^2_{\rm SBS} \sum_{l = 1}^{B_{\rm SBS}} b_{i,l}^{\rm SBS} \frac{1}{F} \sum_{j = 1}^{F} | h_{j, l, u}^{\rm SBS \rightarrow PU} |^2 ~.  
\end{equation}

\subsection{Problem Formulation}


Traditional \ac{CR} assumes that the channels between primary \ac{Rx} and primary and secondary \acp{Tx} (or \ac{Tx} sectors) are known. Sensing whether a particular primary \ac{Tx} (sector) is active, which is a binary detection problem, then allows us to determine whether activating a particular secondary \ac{Tx} is permissible. The key difference in the setup we consider is that the specific set of actively receiving \acp{PU} in the $i$-th timeslot is unknown. This is motivated by the structure of 5G transmissions as discussed in more detail in Sec. \ref{sec: offline_phase}. The beam selection process at the \ac{SBS} is thus based solely on (i) the estimated set of currently active \ac{PBS} beams, $\hat{\mathcal{B}}_i^{\rm PBS}$, and (ii) a probabilistic constraint requiring that, on average, the fraction of \acp{PU} experiencing significant interference remains below a predefined threshold.

Formally, for each interval $i$ and sector $k$, we require that the probability of a randomly selected active \ac{PU} $u \sim \mathcal{P}_k^{\rm PU}$ experiencing SINR below a minimum threshold $\theta$ satisfies 
\begin{align}
    \label{eq: prob_sinr_gen}
   \text{Pr}_{u \sim \mathcal{P}_k^{\rm PU}}&(1\{\text{SINR}_{i,u} < \theta\})  =  \sum_{u_{\text{act}} \in \mathcal{U}_k} \text{Pr}(u = u_{\text{active}})  \\ & \times\text{Pr}\left(1\left\{\text{SINR}_{i,u} < \theta\right\} \mid u = u_{\text{active}}\right) < V ~ \nonumber.
\end{align}
Here, $\theta$ represents the minimum SINR required for reliable \ac{PU} operation, and $V$ denotes the maximum permissible fraction of locations where the SINR may fall below this threshold due to \ac{SBS} transmissions.\footnote{This formulation assumes that, in the absence of \ac{SBS} transmissions, all \acp{PU} meet the SINR requirement. The \ac{PBS} scheduler will typically ensure that this condition is fulfilled. } 
 $\text{Pr}(u = u_{\text{active}}) $ is the spatial probability distribution of the \acp{PU}. We assume in the following that the \ac{SBS} has no knowledge at all about the spatial distribution and thus use a uniform distribution $\text{Pr}(u = u_{\text{active}})={1}/{|\mathcal{U}_k|}$. 

More detailed distributions can, e.g., be learned over time by the \ac{SBS} through uplink transmissions and their association with location-specific channel state information. It can also incorporate probabilistic information about the set of \acp{PU} that might be intended as \acp{Rx} in a given timeslot, e.g., based on traffic patterns and uplink/downlink correlations. These aspects will be explored more in our future work.

Finally, the optimization aims to serve, with a given (or estimated) set of active \ac{PBS} sectors, the maximum number of \acp{SU} without violating the \ac{PU} interference constraint. We furthermore assume that 
the \ac{SBS} throughput is maximized by maximizing the number of \ac{SBS} beams that are active:
\begin{equation}
    \label{eq: problem}
    \max_{b_{i, l}^{\rm SBS}}\sum_{l = 1}^{B_{\rm SBS}} b_{i,l}^{\rm SBS} ~~~
    s.t. \;  (\ref{eq: prob_sinr_gen}), \; \forall i \in \mathcal{I}. 
\end{equation} 

\section{Algorithm} \label{sec: algorithm} 

We now come to the core of this paper, namely the algorithm for determining which \ac{SBS} sectors can transmit in which timeslots. This algorithm proceeds in two phases: an offline phase in which the SBS learns the UE-PBS and UE-SBS channels for each UE location, and an online phase, in which the SBS decides, for each timeslot, which SBS beams may transmit. {\em Notably, in both phases, we do not utilize any information about the active \acp{PU}, including their locations or channel states. } 

\subsection{Offline Phase} \label{sec: offline_phase}
The equations in Sec.~\ref {sec: system_model} assume that the channels between the \ac{PU} and both the \ac{PBS} and the \ac{SBS} are known to the \ac{SBS}—an assumption commonly made in the \ac{CR} literature. However, in typical 5G cellular systems, this assumption does not hold due to the structure of uplink and downlink transmissions. A 5G \ac{BS} might be connected to dozens or hundreds of \acp{UE}, and it is impossible for somebody without access to the scheduler of the \ac{BS} to know which \ac{UE} will be scheduled in the next downlink transmission slot. In particular, downlink transmissions are not necessarily preceded by any uplink activity \cite{dahlman20205g}, \cite[Chap.~32]{molisch2023wireless}. While acknowledgment (ACK) packets are typically sent on the uplink \emph{after} a downlink transmission, this feedback arrives too late to be useful for \ac{CR} applications.\footnote{Even in cases of repeated packet transmissions, such as during streaming—where an ACK packet could suggest future packets might be sent to the same UE—these ACKs may occur at time-frequency resources with no discernible relation (from the \ac{SBS} perspective) to the preceding downlink signal. Moreover, they may be indistinguishable from UE-initiated transmissions, making them ineffective for inferring the location or channel of the \ac{PU}.} Furthermore, in multicast scenarios, uplink transmissions may be absent altogether, rendering it even more difficult to identify the active \ac{PU} locations or channel conditions.

We propose, however, that what {\em can} be achieved is knowledge of the \ac{PBS}-\ac{PU} and \ac{SBS}-\ac{PU} channels (more precisely, between each sector of \ac{PBS} and \ac{PU}, and similarly for the \ac{SBS}) for a given UE location. To this end, we introduce the following protocol:
\begin{enumerate}[leftmargin=*]
    \item A \ac{SU} observes the transmission from the \ac{PBS} and records its time of observation, primary signal characteristics (e.g., occupied frequency), as well as its own location. 
    \item Within a short time (e.g., at the next secondary uplink transmission opportunity), the \ac{SU} transmits this information to the \ac{SBS}. This transmission provides the \ac{SBS} not only with the observations made by the \ac{SU}, but also enables it to measure the channel from the \ac{SU} location to the \ac{SBS}, which—due to channel reciprocity\footnote{In frequency division duplex systems, the small-scale-averaged characteristics are reciprocal, while in time division duplex systems, even complex channel gains might be reciprocal.}—is also the channel from the \ac{SBS} to the UE. Furthermore, the \ac{SBS} continuously monitors transmissions from the \ac{PBS} and records the corresponding information along with timestamps.\footnote{Note that the \ac{SBS} does not require knowledge of the \ac{PBS} geometry or its beam patterns; it can describe active \ac{PBS} beams and infer user associations solely based on the observed channel states between the \ac{PBS} and the \ac{SBS}.}
    \item The \ac{SBS} thus obtains knowledge of the \ac{PBS} transmission status, the UE's location, the  \ac{PBS} - UE channel (pathloss), and the \ac{SBS} - UE channel. While the sensing UE is an \ac{SU}, the channel and the received power from a BS to a UE are independent of whether the UE is a \ac{PU} or an \ac{SU}!
\end{enumerate}
These channels need to be determined at every \ac{UE} location\footnote{Since we only need the small-scale averaged (SSA) channel characteristics, it is sufficient to measure at one location within a "stationarity" region" of the channel, which is typically on the order of a few tens of meters in outdoor environments, \cite[Chapter 7]{molisch2023wireless}} within the cell; thus the observing \acp{SU} have to cover all locations within the cell and observe any combinations of \ac{PBS} beam transmission. This leads to a long required duration for the learning process; the corresponding channel between a UE in a particular location and a BS is assumed to be time invariant during the learning process, which is approximately fulfilled since dominant scatterers like buildings are static. Residual variations, e.g., due to moving scatterers like cars, are neglected here because they typically have a small radar cross section and only introduce minor channel variations; their impact will be considered in future work. 

The output of the offline phase are thus the channels for different \ac{PBS} beam combinations, $\mathcal{B} \triangleq \{b_k\}_{k = 1}^{B_{\rm PBS}} \in \{0,1\}^{B_{\rm PBS}}$, and the corresponding expected energy label vector $g(\mathcal{B}) \in \mathbb{R}^{B_{\rm SBS}}$ by averaging over many frames. Assuming Gaussianity of the noise-induced variations, the $l^{\text{th}}$ entry of $g(\mathcal{B})$ is given by 
\begin{equation}
g_l(\mathcal{B}) = \sum_{k = 1}^{B_{\rm PBS}}\sum_{j = 1}^{F} b_k^{\rm PBS} \sigma^2_{\rm PBS} \left| h_{j,k,l}^{\rm PBS \rightarrow SBS} \right|^2 + \sigma^2_{\rm noise}.
\end{equation}
This allows computation of an SINR table for each location $\boldsymbol{z}_u$, $\text{SINR}_{u}$ with respect to $\mathcal{B}^{\rm PBS}$ and $\mathcal{B}^{\rm SBS}$, {where each entry corresponds to a unique combination of \ac{PBS} and \ac{SBS} beam activity. The index \( i \) is omitted, as the table remains the same in all sensing intervals and does not depend on \( i \).}
\subsection{Online Phase}
In each time interval $i$ the \ac{SBS} needs to (i) estimate, at the end of the sensing phase, the active \ac{PBS} beams $\hat{\mathcal{B}}_i^{\rm PBS}$ based on the $N$ collected samples, and (ii) select a subset of \ac{SBS} beams to transmit packets to the \acp{SU}, ensuring that the service quality of the \acp{PU} $\text{SINR}_{i,u}$ for the given $b_{i,k}^{\rm PBS}$ and the chosen $b_{i,l}^{\rm SBS}$ is not excessively degraded as per (\ref{eq: prob_sinr_gen}).

\subsubsection{Detection of PBS Beams}
During the offline phase, we assume that the \ac{SBS} can perfectly associate \ac{PBS} activity with the received signals, owing to the static nature of the environment and the extended measurement duration. In contrast, during the online phase, the \ac{SBS} must detect \ac{PBS} activity based on a limited number of sensing samples, which may introduce detection inaccuracies. To detect the active (ON) beams of the \ac{PBS}, we employ a simple energy detector based on a maximum-likelihood (ML) rule. Specifically, given the observed energies across \ac{SBS} beams, we select the most likely configuration among the $2^{B_{\rm PBS}}$ possible \ac{PBS} beam combinations. The energy measured during sensing interval $i$ on beam $l$ of the \ac{SBS} is defined as 
\begin{equation}
v_{i,l}^{\rm SBS} \triangleq \frac{1}{N}\sum_{n = 1}^{N} \sum_{j = 1}^{F} |r^{\rm SBS}_{i, j, n, l}|^2.
\end{equation}We then stack the energy across all beams into a feature vector $\boldsymbol{x}_i \triangleq \left[ v_{i,1}^{\rm SBS}, \dots, v_{i,B_{\rm SBS}}^{\rm SBS} \right]$, where $\boldsymbol{x}_i \in \mathbb{R}^{B_{\rm SBS}}$.
Given the observed energy $\boldsymbol{x}_i$, we estimate the active \ac{PBS} beams via maximum-likelihood detection as $\hat{\mathcal{B}}_i^{\rm PBS} = \arg\min_{\mathcal{B}} \lVert \boldsymbol{x}_i - g(\mathcal{B}) \rVert^2$. 
\subsubsection{SBS Transmission Decision}
Based on the identified active transmission beams of the \ac{PBS}, we assess whether, and in which directions, the \ac{SBS} is permitted to transmit without catastrophic interference to actively receiving \acp{PU}; i.e., solve (\ref{eq: problem}). Taking the SINR table entries for the estimated $\hat{\mathcal{B}}_i^{\rm PBS}$, we can find $\mathcal{B}_i^{\rm SBS} = \arg\max_{\mathcal{B}} \sum_{l = 1}^{B_{\rm SBS}} b_l$ subject to the constraint that the average SINR for \acp{PU} in $\mathcal{U}_k$ remains above a given threshold under the selected configuration, as formalized in Eq. (\ref{eq: prob_sinr_gen}). In other words, we search for all the \ac{SBS} beam combinations that do not lead to catastrophic interference, i.e., ensure that the increase in \acp{PU} whose SINR threshold exceeds $\theta$ increases beyond a threshold $V$. We note that besides this constraint, other criteria, like maximum reduction of average capacity or outage capacity, could be imposed. Furthermore, the maximization of the number of secondary beams could be replaced by maximizing throughput for the scheduled \acp{SU} (since the \ac{SBS} knows the channels to the \acp{SU}) could be employed. While all these results give slightly different optimization results, the protocols and algorithmic frameworks remain unchanged.

\section{Numerical Evaluation} \label{sec: numerical_eval}
\textbf{Dataset}. We evaluate our algorithm by using ray tracing to obtain realistic propagation channels; we use the \emph{Wireless InSite} tool for generating \acp{MPC} parameters (see (\ref{eq: bs_channel})) between the \ac{PBS} and \ac{SBS}, as well as between each (single-antenna) \ac{UE} and the \acp{BS}. Simulations shown in the following are done on the USC campus near the TCC building, at a carrier frequency of 2.5 GHz, bandwidth of 1 MHz, and using beamforming via a spatial Fourier transformation of a uniform linear array as commonly used in 5G \cite{molisch2023wireless}. Further simulation details are omitted due to space constraints.

The \ac{PBS} does not perform power control; its transmit power is chosen such that it provides a minimum of 3 dB \ac{SNR} to each \ac{PU} in the environment.
In our setup, we found $\sigma^2_{\rm PBS} = 5$~dBm. The \ac{SBS} is not assumed to have power control capability, but its transmit power $\sigma^2_{\rm SBS}$ is a system parameter we vary in the subsequent simulations to study its impact. 
We further assume that the beamforming gains of both \ac{PBS} and \ac{SBS} are absorbed into the transmit powers; thus, $\sigma^2_{\rm PBS}$ and $\sigma^2_{\rm SBS}$ represent the equivalent isotropically radiated power (EIRP). The activity of each \ac{PBS} beam $b_{i,k}$ is modeled as an i.i.d. Bernoulli random variable where $b_{i,k} = 1$ with probability $1/2$.

\textbf{Baselines}. \emph{Multi-Detection Binary Access}, which uses the same hardware setup with directional antennas. The \ac{SBS} predicts the individual beam activity, $\hat{\mathcal{B}}_i^{\rm PBS}$, and the transmission is binary—\ac{SBS} either transmits all of the beams (if the SINR constraints are satisfied) or remains silent. \emph{Binary-Detection Binary Access}, a conventional \ac{SS} method, detects only the presence of any \ac{PBS} activity without directionality. It transmits omnidirectionally if no \ac{PBS} beam is active (unlike the former, which can still transmit even if there is \ac{PBS} activity, as long as the SINR constraints are satisfied); otherwise, it remains silent. For consistency, the transmission power levels are kept identical to those in the directional case.
\subsection{Results}
We evaluate our system using three criteria defined as follows.

\textbf{Probability of Missed Opportunity (PMO)}. This metric quantifies the percentage of \ac{SBS} beams that satisfy the SINR constraints and could have been used for transmission, but were not selected, primarily due to misdetection of active \ac{PBS} beams. We define the total number of admissible transmissions as $B_{\rm mo}\triangleq \sum_{i = 1}^I \sum_{l = 1}^{B^{\rm SBS}} b_{i, l}^{\text{SBS, GT}}$, where $\text{GT}$ denotes the ground truth obtained under the assumption of perfect knowledge of the \ac{PBS} beam activity in each interval. Accordingly, the PMO is defined as
\begin{equation}
    C_{\rm mo} \triangleq \frac{1}{B_{\rm mo}}\sum_{i = 1}^I\sum_{l = 1}^{B_{\rm SBS}} 1\{b_{i, l}^{\text{SBS, GT}} = 1\} 1\{{b}_{i, l}^{\rm SBS} = 0\}.
\end{equation}

\textbf{Probability of Catastrophic Interference (PCI)}. The main constraint for the secondary system is the potential catastrophic interference that PUs experience once the secondary system is active. $C_{\rm int}$ describes the percentage of the PUs that had severe interference after the SBS transmission, where the total number of beams active in the environment is defined as $B_{\rm int} \triangleq\sum_{i = 1}^I\sum_{l = 1}^{B_{\rm PBS}}b_{i, k}^{\text{PBS}}$. Then, the PCI is defined as  
\begin{equation}C_{\rm int} \triangleq \frac{1}{B_{\rm int}}\sum_{i = 1}^I\sum_{k =1}^{B_{\rm PBS}}1\{(\frac{b_{i,k}}{|\mathcal{U}_k|}\sum_{u \in \mathcal{U}_k}1\left\{\text{SINR}_{i,u}<\theta\right\}) > V\}.
\end{equation}
\\~\\
\textbf{Throughput}. Since the main objective of all \ac{CR} efforts is to serve the secondary users, we assess the performance of the proposed scheme in terms of the throughput of the secondary system, which is evaluated as 
\begin{equation}
    C_{\rm thru}   \triangleq \frac{1}{I}\sum_{i = 1}^I\sum_{u =1 }^U\frac{1\{\text{SINR}_{i,u}^{\rm SBS}>\theta\}}{U}.
\end{equation}

Secondly, the $\text{SINR}^{\rm SBS}_{i,u}$ is computed based on the SBS beam configuration, analogous to the SINR calculation for the \ac{PBS}, but with the PBS and SBS roles interchanged, as the evaluation is for the \acp{SU}.
\begin{figure}[h!]
    \centering

    \includegraphics[trim = {0, 0, 0, 1cm}, clip, width=1\columnwidth]{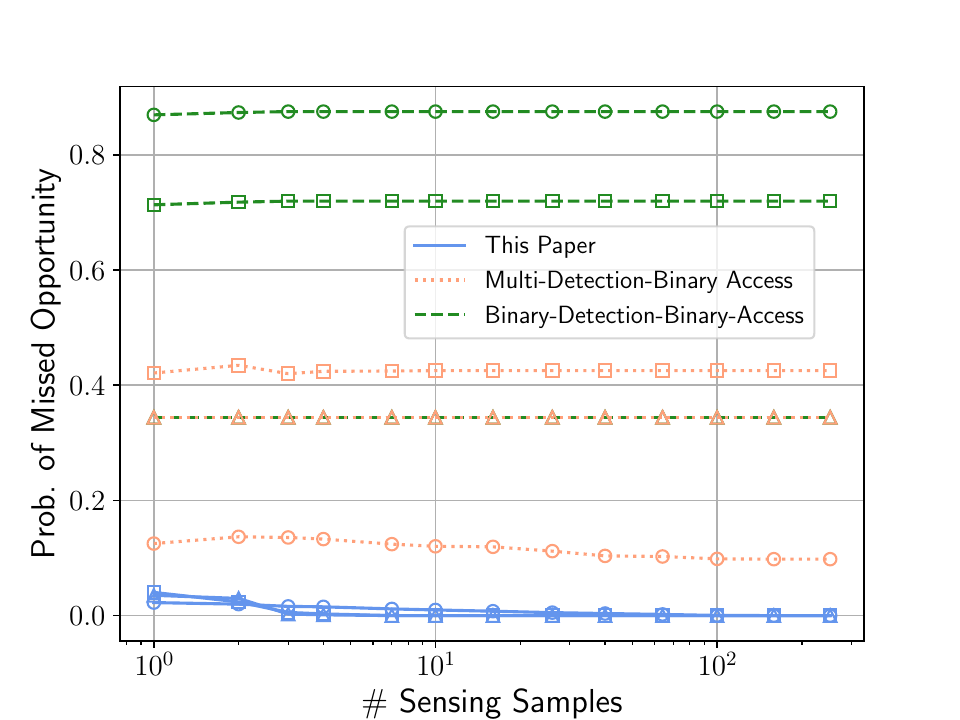}

    \caption{PMO Comparison for different $\sigma^2_{\rm SBS} \backslash  \sigma^2_{\rm PBS}$ as {\color{gray}$\circ$}: -20 dB, {\color{gray}$\square$}:  -5 dB, {\color{gray}$\triangle$}: 5 dB. MDBA and BDBA for 5 dB overlap.}
    \label{fig: missed}
\end{figure}

In Fig. (\ref{fig: missed}), we compare the PMO as a function of the number of sensing samples that SBS gathered and show results for different SBS powers. In the case of perfect estimation of the PBS beam state, our method, by definition, incurs no missed transmission opportunities; this is achieved after 100 samples. However, when the PBS beam state is misestimated, the \ac{SBS} may refrain from transmitting in certain directions due to erroneously predicted interference. The binary-detection baseline, which waits for complete PBS silence, misses the most transmission opportunities. Although the multi-detection baseline senses directionally, it still transmits only when all directions are deemed safe, leading to more missed opportunities than our method. We also see that while the PMO for our method depends mostly on the number of sensing samples, and thus on the accuracy of the \ac{PBS} beam identification, the PMO for the other methods is dominated by the ratio of SBS to PBS power. This can be explained by the increased probability that even secondary beams with small power coupling into PBS beams create significant interference, which, according to the principle of the baseline, prevents {\em all} transmissions from the SBS and thus increases the PMO. 
\begin{figure}[h!]
    \centering

    \includegraphics[trim = {0, 0, 0, 1cm}, clip, width=1\columnwidth]{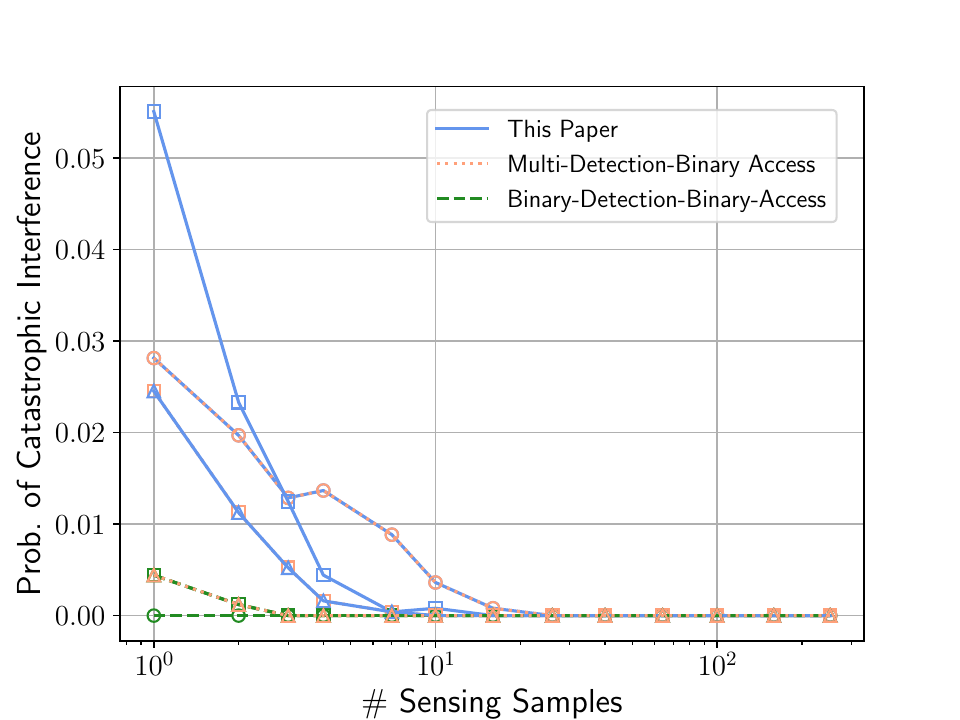}

    \caption{PCI Comparison, for different $\sigma^2_{\rm SBS} \backslash  \sigma^2_{\rm PBS}$ as {\color{gray}$\circ$}: -20 dB, {\color{gray}$\square$}: - 5 dB, {\color{gray}$\triangle$}: 5 dB. Our method and MDBA for -20 dB overlap. MDBA (5 dB) and BDBA (-5 dB) overlap. Our method (5 dB) and MDBA (-5 dB) overlap.}
    \label{fig: lethal}
\end{figure}

\begin{figure}[h!]
    \centering
    \includegraphics[trim = {0, 0, 0, 1cm}, clip, width=1\columnwidth]{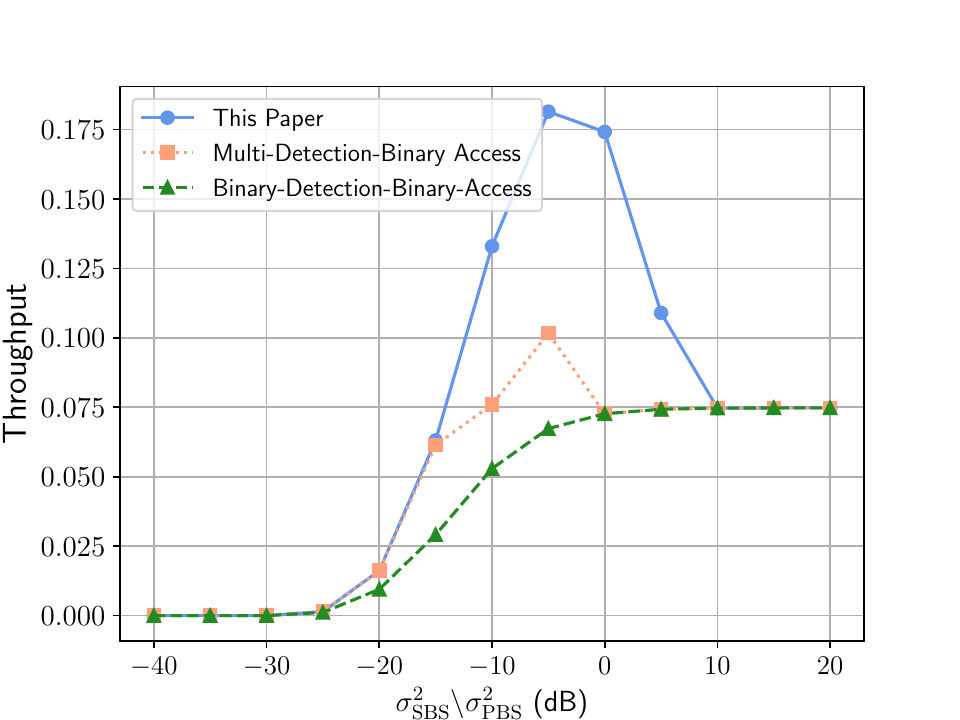}
    \caption{Throughput for the Secondary Users}
    \label{fig: thru}
\end{figure}

A similar study is done for the PCI in Fig. (\ref{fig: lethal}). Note, however, that absence of CI does {\em not} mean that there are zero \acp{PU} impacted by interference, just that the percentage of catastrophically interfered \acp{PU} in one beam remains below the threshold $V$. Note that for the low number of sensing samples, our algorithm may lead to higher PCI, since it exploits {\em all} apparent transmission opportunities (even when they are just a consequence of the erroneous PBS beam activity). However, the use of a longer sensing window (100 samples) essentially eliminates this problem. While this reduces the available transmission time for the secondary payload, this is more than compensated by the increased probability of transmission 
as shown in Fig.~(\ref{fig: thru}).

Finally, we analyze the throughput of the \ac{SBS} as a function of the power ratio SBS/PBS, as shown in Fig. (\ref{fig: thru}). We observe that throughput increases with \ac{SBS} power up to \(-5\,\text{dB}\). At low power levels, the SBS transmits frequently but covers only a small portion of its intended area (assuming the SBS cell size equals that of the PBS). In this regime, our strategy always allows transmission in all directions, effectively reducing to the \emph{Multi-Detection Binary Access} method. Between \(-10\,\text{dB}\) and \(-5\,\text{dB}\), however, our method can exploit diverse beam combinations not used by the baselines, achieving higher throughput. Beyond \(-5\,\text{dB}\), interference becomes significant, limiting SBS activity. At this point, the SBS must wait for complete PBS silence, causing all methods to converge to the \emph{Binary-Detection Binary Access} strategy. Thus, careful tuning of SBS power is essential to maximize throughput while keeping catastrophic interference within acceptable limits. Further improvements can be achieved through the use of adaptive power control, which will be investigated in future work. 

\section{Conclusion} \label{sec: conclusion}

We considered a realistic infrastructure-based system with multi-beam antennas at both primary and secondary base stations. We determined the set of beams from the secondary base station that allowed secondary transmissions into different beam directions without creating excess interference for the PUs in the environment. We leveraged and showed the results for different power levels of the SBS. Our findings underscore the potential of carefully optimized multi-beam configurations to enhance coexistence and efficiency in shared spectrum.

In our future work, we will pursue the following directions: (i) relax the offline assumption of perfect channel statistics and replace it by robust estimation under partial observations; (ii) incorporate per-beam power control at both primary and secondary base stations; (iii) enhance scalability with low-complexity heuristics that offer provable approximation and regret guarantees; (iv) evaluate robustness to sensing errors, mobility, and noise-induced channel-estimation errors using stochastic constraints and reliability analyses; (v) exploit temporal correlation in beam activity to improve low-SNR detection; and (vi) develop ML-based SINR predictors to reduce data-collection overhead.

\section*{ACKNOWLEDGEMENT} 
The authors greatly appreciate the constructive discussions with Dr. Naveed Ahmed Abbasi.
The work was financially
supported by NSF Grant 2229535.

\bibliographystyle{IEEEtran}
\bibliography{bibli}

\end{document}